\documentclass{ws-ijmpd}
\usepackage[super,compress]{cite}
\begin{document}

\markboth{Peter A. Hogan, Dirk Puetzfeld}
{Gravitational Radiation from an Accelerating Massive Particle in General Relativity}

%
\catchline{}{}{}{}{}
%

\title{GRAVITATIONAL RADIATION FROM AN ACCELERATING MASSIVE PARTICLE IN GENERAL RELATIVITY}

\author{PETER A. HOGAN}

\address{School of Physics, University College Dublin \\
         Belfield, Dublin 4, Ireland\\
         peter.hogan@ucd.ie}

\author{DIRK PUETZFELD}

\address{University of Bremen, Center of Applied Space Technology and Microgravity (ZARM)\\
        28359 Bremen, Germany\\
        and\\
        School of Physics, University College Dublin \\
        Belfield, Dublin 4, Ireland\\
        dirk@puetzfeld.org}

\maketitle

\begin{history}
\received{Day Month Year}
\revised{Day Month Year}
\end{history}

\begin{abstract}
A comprehensive description is given of a space--time model of an accelerating massive particle. The particle radiates gravitational waves with optical shear. The wave fronts are smoothly deformed spheres and the particle experiences radiation reaction, similar to an accelerating charged particle, and a loss of mass described by a Bondi mass--loss formula. The space--time is one of the Bondi--Sachs forms but presented in a form here which is particularly suited to the construction of the model particle. All details of the calculations are given. A detailed examination of the gravitational field of the particle is provided which illustrates the presence of gravitational radiation and also exhibits, in the form of a type of singularity found in some Robinson--Trautman space--times, the absence of an external field to supply energy to the particle.
\end{abstract}


\ccode{PACS numbers: 04.20.-q; 04.30.-w; 04.20.Cv}


\section{Introduction}\label{sec:1}

The papers by Bondi et al.\ \cite{Bondi:etal:1962} and Sachs \cite{Sachs:1962} on gravitational radiation from isolated sources are classics in the general relativity literature. Notable spin--offs include the Bondi--Metzner--Sachs (BMS) group (see Ref.\ \refcite{Alessio:Esposito:2018} for a recent pedagogical review) and recent studies of the Bondi mass--loss formula when a positive cosmological constant is present (for example Refs.\ \refcite{Winicour:2012,Kubeka:Bishop:2014,Saw:2016,Saw:2018,Giannakopoulos:etal:2022} and references therein). The Bondi--Sachs approach has also been used to study the geodesic hypothesis in general relativity \cite{Hogan:Robinson:1986}{}. It is to this we turn in the present paper to provide extensive details of a version of the Bondi--Sachs method which is particularly suited to this project and also to illustrate the appearance of directional singularities in the gravitational field of an accelerating radiating massive particle when an external field is absent. This is notwithstanding the fact that in our model of the massive particle the wave fronts of the radiation produced by the particle are smoothly deformed spheres (have no singular points). To ensure that the paper is self contained we describe in detail in section \ref{sec:2} the version of the Bondi--Sachs method which we use. The model accelerating massive particle is constructed in section \ref{sec:3} and its gravitational field is analyzed in section \ref{sec:4}. The paper ends with a brief discussion in section \ref{sec:5}.

\section{The Basic Space--Time}\label{sec:2}

Our space--time model of an accelerating massive particle generating gravitational radiation will be a special case of the space--time constructed by Bondi et al.\cite{Bondi:etal:1962}{}, in the case of axial symmetry, and more generally by Sachs \cite{Sachs:1962} and Newman and Unti \cite{Newman:Unti:1962}{}. We utilize here, and describe in detail, a version of this space--time given in Ref.\ \refcite{Hogan:Trautman:1987} which is particularly useful for our purposes. In all of these space--times the gravitational waves envisaged are simple in the sense that they have easily identifiable wave fronts. This means that the histories of the wave fronts in space--time are null hypersurfaces. The first consideration in building a local coordinate system based on these null hypersurfaces, in which to express the line element of the space--time, is to take a light--like vector field $k$ which, in a general local coordinate system $\{x^i\}$ with $i=1, 2, 3, 4$, satisfies
\begin{equation}\label{2.1}
k=k^i\frac{\partial}{\partial x^i}\ \ \ {\rm with}\ \ \ k^i=g^{ij}\,u_{,j}\ \ \ {\rm and}\ \ \ g_{ij}\,k^i\,k^j=0\ .
\end{equation}
Here $g_{ij}=g_{ji}$ are the components of the metric tensor in the coordinates $\{x^i\}$ and $g^{ij}$ is the inverse of the metric tensor defined by $g^{ij}\,g_{jk}=\delta^i_k$. Also $u(x^i)$ is a differentiable function of the coordinates and the comma denotes partial differentiation with respect to $x^j$. Thus $u(x^i)={\rm constant}$, constitutes a family of null hypersurfaces in the space--time. Let $r$ be an affine parameter along the integral curves of the vector field $k$. It follows from (\ref{2.1}) that these curves are null geodesics and are the generators of the null hypersurfaces. We can make use of $u, r$ as coordinates so that if we choose as local coordinates $x^i=(x^A, r, u)$ for $i=1, 2, 3, 4$ and $A=1, 2$ then the line element of the space--time in these coordinates, as a result of (\ref{2.1}), takes the form
\begin{equation}\label{2.2}
ds^2=-h_{AB}\,(dx^A+a^A\,du)(dx^B+a^B\,du)+2\,du\,dr+c\,du^2\ ,
\end{equation}
where $h_{AB}=h_{BA}, a^A$ and $c$ are six functions of the coordinates $x^i=(x^A, r, u)$. Proceeding as in Ref.\ \refcite{Hogan:Trautman:1987} we write $h_{AB}=P^2\gamma_{AB}$ with $P$ chosen so the $\det(\gamma_{AB})=1$ and then parametrize $(\gamma_{AB})$ as
\begin{equation}\label{2.3}
(\gamma_{AB})=\left (\begin{matrix}e^{2\alpha}\cosh2\beta&\sinh2\beta\\
\sinh2\beta&e^{-2\,\alpha}\cosh2\beta\\\end{matrix}\right )\ ,
\end{equation}
with $P, \alpha, \beta$ functions of $(x^A, r, u)$. It will be useful later to write $P=r\,p^{-1}$ with $p=p(x^A, r, u)$. If for simplicity of notation we write $x^1=x$ and $x^2=y$ and in addition define, in place of $a^1$ and $a^2$,
\begin{equation}\label{2.4}
a=a^1e^{\alpha}\cosh\beta+a^2e^{-\alpha}\sinh\beta\ ,
\end{equation}
and
\begin{equation}\label{2.5}
b=a^1e^{\alpha}\sinh\beta+a^2e^{-\alpha}\cosh\beta\ ,
\end{equation}
then the line element (\ref{2.2}) takes the final form
\begin{eqnarray}
ds^2&=&2\,dr\,du+c\,du^2-r^2p^{-2}\{(e^{\alpha}\cosh\beta\,dx+e^{-\alpha}\sinh\beta\,dy+a\,du)^2\nonumber\\
&&+(e^{\alpha}\sinh\beta\,dx+e^{-\alpha}\cosh\beta\,dy+b\,du)^2\}\ .\label{2.6}
\end{eqnarray}
We note that the six functions of the four coordinates in the line element (\ref{2.2}) have now been replaced by the six functions $p, \alpha, \beta, a, b, c$ of $(x, y, r, u)$. The light--like vector field $k=\partial/\partial r$ and the null geodesic integral curves of $k$ have expansion and shear given by
\begin{equation}\label{2.7}
\rho=\frac{\partial}{\partial r}(r\,p^{-1})\ \ \ {\rm and}\ \ \ \sigma=\cosh2\beta\frac{\partial\alpha}{\partial r}+i\,\frac{\partial\beta}{\partial r}\ ,
\end{equation}
respectively.

\begin{figure}[pb]
    \centerline{\includegraphics[width=0.4\textwidth]{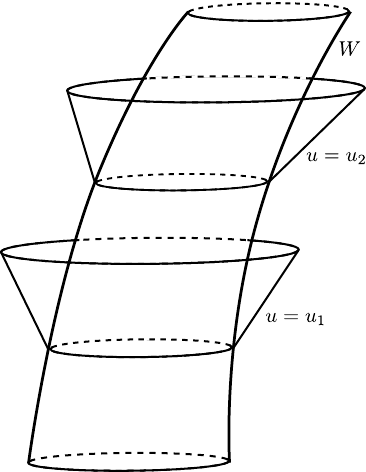}}
    \vspace*{8pt}
    \caption{\label{fig_1} $W$ is a time--like world tube enclosing the history in space-time of an isolated gravitating source. The null hypersurfaces $u={\rm constant}$ ($u=u_1$ and $u=u_2$ being examples) are envelopes of future null cones with vertices on $W$. The null hypersurfaces are generated by null geodesics with $r$ an affine parameter along them. The null geodesics are labelled by pairs of stereographic coordinates $x, y$.}
\end{figure}
    
For our purposes later we shall assume that the six functions can each be expressed as a power series in powers of $r$ with coefficients functions of $x, y, u$ as follows:
\begin{eqnarray}
p&=&p_0\left (1+\frac{q_2}{r^2}+O\left (\frac{1}{r^4}\right )\right )\ ,\label{2.8}\\
\alpha&=&\frac{\alpha_1}{r}+\frac{\alpha_3}{r^3}+O\left (\frac{1}{r^4}\right )\ ,\label{2.9}\\
\beta&=&\frac{\beta_1}{r}+\frac{\beta_3}{r^3}+O\left (\frac{1}{r^4}\right )\ ,\label{2.10}\\
a&=&\frac{a_2}{r^2}+\frac{a_3}{r^3}+O\left (\frac{1}{r^4}\right )\ ,\label{2.11}\\
b&=&\frac{b_2}{r^2}+\frac{b_3}{r^3}+O\left (\frac{1}{r^4}\right )\ ,\label{2.12}\\
c&=&r\,c_{-1}+c_0+\frac{c_1}{r}+\frac{c_2}{r^2}+O\left (\frac{1}{r^3}\right )\ .\label{2.13}
\end{eqnarray}
These expansions have been determined in Ref.\ \refcite{Hogan:Trautman:1987} by utilizing some of Einstein's vacuum field equations and simplifying using allowable coordinate transformations which preserve the form of the line element (\ref{2.6}). With these series assumptions the expansion and shear in (\ref{2.7}) of the vector field $k$ become
\begin{equation}\label{2.14}
\rho=\frac{1}{r}+\frac{2\,q_2}{r^3}+O\left (\frac{1}{r^5}\right )\ ,
\end{equation}
and 
\begin{equation}\label{2.15}
    \sigma=-\frac{(\alpha_1+i\,\beta_1)}{r^2}-\frac{\{2\,\alpha_1\,\beta_1^2+3\,(\alpha_3+i\,\beta_3)\}}{r^4}+O\left (\frac{1}{r^5}\right )\ ,
\end{equation}
respectively.

We will now impose the vacuum field equations in order to obtain further information on the coefficients of the powers of $r$ appearing explicitly in (\ref{2.8})--(\ref{2.13}). We gradually work through Einstein's vacuum field equations $R_{ij}=0$ with $R_{ij}=R_{ji}$ the components of the Ricci tensor in coordinates $x^i=(x, y, r, u)$ with $i=1, 2, 3, 4$ calculated with the metric tensor given by (\ref{2.6}) and (\ref{2.8})--(\ref{2.12}). The component $R_{11}$ has a term linear in $r$, a term independent of $r$, an $r^{-1}$--term, an $r^{-2}$--term and so on. The first three terms here, when equated to zero, give
\begin{eqnarray}
c_{-1}&=&-2\,H=-2\,\frac{\partial}{\partial u}(\log p_0)\ ,\nonumber \\
c_0&=&p_0^2\left (\frac{\partial^2}{\partial x^2}+\frac{\partial^2}{\partial y^2}\right )\log p_0\equiv\Delta\log p_0\ ,\label{2.16}
\end{eqnarray}
and
\begin{equation}\label{2.17}
\frac{\partial A}{\partial x}+\frac{\partial B}{\partial y}=0\ ,
\end{equation}
with
\begin{eqnarray}
A&=&p_0^{-2}a_2-p_0^2\left (\frac{\partial}{\partial x}(p_0^{-2}\alpha_1)+\frac{\partial}{\partial y}(p_0^{-2}\beta_1)\right )\ ,\label{2.18}\\
B&=&p_0^{-2}b_2-p_0^2\left (\frac{\partial}{\partial x}(p_0^{-2}\beta_1)-\frac{\partial}{\partial y}(p_0^{-2}\alpha_1)\right )\ .\label{2.19}
\end{eqnarray}
Hence we have $R_{11}=O(r^{-2})$ and we will refine this later. With these conditions holding we find that $R_{22}=O(r^{-2})$ and $R_{12}=O(r^{-2})$ automatically. Next
\begin{equation}\label{2.20}
R_{33}=O\left (\frac{1}{r^6}\right )\ \ \Rightarrow\ \ q_2=\frac{1}{2}(\alpha_1^2+\beta_1^2)\ ,
\end{equation}
and now
\begin{equation}\label{2.21}
R_{34}=O\left (\frac{1}{r^4}\right )\ \ \ {\rm and}\ \ \ R_{44}=O\left (\frac{1}{r^2}\right )\ .
\end{equation}
Also
\begin{equation}\label{2.22}
R_{13}=O\left (\frac{1}{r^3}\right )\ \ \Rightarrow\ \ a_2=p_0^4\left (\frac{\partial}{\partial x}(p_0^{-2}\alpha_1)+\frac{\partial}{\partial y}(p_0^{-2}\beta_1)\right )\ ,
\end{equation}
and
\begin{equation}\label{2.23}
R_{23}=O\left (\frac{1}{r^3}\right )\ \ \Rightarrow\ \ b_2=p_0^4\left (\frac{\partial}{\partial x}(p_0^{-2}\beta_1)-\frac{\partial}{\partial y}(p_0^{-2}\alpha_1)\right )\ ,
\end{equation}
so that $A$ and $B$ in (\ref{2.18}) and (\ref{2.19}) vanish. Also
\begin{equation}\label{2.24}
R_{14}=O\left (\frac{1}{r^2}\right )\ \ \Rightarrow\ \ \frac{\partial A}{\partial u}=0\ ,
\end{equation}
with $A$ given by (\ref{2.18}) and 
\begin{equation}\label{2.25}
R_{24}=O\left (\frac{1}{r^2}\right )\ \ \Rightarrow\ \ \frac{\partial B}{\partial u}=0\ ,
\end{equation}
with $B$ given by (\ref{2.19}). We can summarize the consequences of the vacuum field equations at this point as follows: $c_{-1}$ (and therefore the function $H$) and $c_0$ are given by (\ref{2.16}), $q_2$ is given by (\ref{2.20}) while $a_2$ and $b_2$ are given by (\ref{2.22}) and (\ref{2.23}). This results in the following expressions for the Ricci tensor components:
\begin{eqnarray}
R_{AB}&=&O\left (\frac{1}{r^2}\right )\ ,\ R_{A4}=O\left (\frac{1}{r^2}\right )\ ,\ R_{44}=O\left (\frac{1}{r^2}\right )\ ,\nonumber\\
R_{A3}&=&O\left (\frac{1}{r^3}\right )\ ,\ R_{34}=O\left (\frac{1}{r^4}\right )\ ,\ R_{33}=O\left (\frac{1}{r^6}\right )\ . \label{2.26}
\end{eqnarray} 
We must now refine the estimates on the right hand sides of these equations in order to obtain further useful information on the coefficients in the expansions (\ref{2.8})--(\ref{2.13}).

If we write $c_1=-2\,m(x, y, u)$ (this provocative notation will be a useful guide later) we find that requiring $R_{44}=O(r^{-3})$  
\begin{eqnarray}
&&-2\,\dot m+6\,m\,H+\frac{1}{2}\Delta c_0-3\,H\,p_0^2\frac{\partial}{\partial y}(p_0^{-2}b_2) -3\,H\,p_0^2\frac{\partial}{\partial x}(p_0^{-2}a_2)-6\,H\,\dot q_2+2\,\frac{\partial^2q_2}{\partial u^2}\nonumber\\
&&+2\,\alpha_1\left\{p_0^2\frac{\partial^2H}{\partial x^2}-p_0^2\frac{\partial^2H}{\partial y^2}\right\}+4\,\beta_1\,p_0^2\frac{\partial^2H}{\partial x\partial y}-2\,(\dot\alpha_1^2+\dot\beta_1^2)+\frac{\partial}{\partial u}\left (p_0^2\frac{\partial}{\partial y}(p_0^{-2}b_2)\right )\nonumber\\
&&+\frac{\partial}{\partial u}\left (p_0^2\frac{\partial}{\partial x}(p_0^{-2}a_2)\right )-4\,q_2\,\dot H+8\,q_2\,H^2+4\,\frac{\partial H}{\partial y}\left (p_0\,\frac{\partial p_0}{\partial x}\,\beta_1-p_0\,\frac{\partial p_0}{\partial y}\,\alpha_1\right )\nonumber \\
&&+4\,\frac{\partial H}{\partial x}\left (p_0\,\frac{\partial p_0}{\partial x}\,\alpha_1+p_0\,\frac{\partial p_0}{\partial y}\,\beta_1\right )=0\ ,\label{2.27}
\end{eqnarray}
with a dot here and throughout indicating differentiation with respect to $u$ and the function $H$ and the differential operator $\Delta$ defined in (\ref{2.16}). The coefficients of $\alpha_1$ and $\beta_1$ here can be simplified so that this equation reads
\begin{eqnarray}
&&-2\,\dot m+6\,m\,H+\frac{1}{2}\Delta c_0-2\,(\dot\alpha_1^2+\dot\beta_1^2)-4\,q_2\,\dot H+8\,q_2\,H^2-6\,H\,\dot q_2\nonumber \\
&&+2\,\alpha_1\left\{\frac{\partial}{\partial x}\left (p_0^2\frac{\partial H}{\partial x}\right )-\frac{\partial}{\partial y}\left (p_0^2\frac{\partial H}{\partial y}\right )\right\}+2\,\beta_1\left\{\frac{\partial}{\partial x}\left (p_0^2\frac{\partial H}{\partial y}\right )+\frac{\partial}{\partial y}\left (p_0^2\frac{\partial H}{\partial x}\right )\right\}\nonumber\\
&&+\frac{\partial}{\partial u}\left\{p_0^2\left (\frac{\partial}{\partial x}(p_0^{-2}a_2)+\frac{\partial}{\partial y}(p_0^{-2}b_2)\right )+2\,\frac{\partial q_2}{\partial u}\right\}\nonumber\\
&&-3\,H\,p_0^2\left (\frac{\partial}{\partial x}(p_0^{-2}a_2)+\frac{\partial}{\partial y}(p_0^{-2}b_2)\right )=0.\label{2.28}
\end{eqnarray}
Writing
\begin{equation}\label{2.29}
-4\,q_2\,\dot H=\frac{\partial}{\partial u}(-4\,q_2\,H)+4\,\dot q_2\,H\ ,
\end{equation}
and introducing the variable $M(x, y, u)$ defined by 
\begin{eqnarray}
M(x, y, u)&=&m(x, y, u)-\dot q_2+2\,q_2\,H-\frac{1}{2}p_0^2\left\{\frac{\partial}{\partial x}(p_0^{-2}a_2)+\frac{\partial}{\partial y}(p_0^{-2}b_2)\right\},\label{2.30}
\end{eqnarray}
we can put (\ref{2.28}) in the form
\begin{eqnarray}
&&-2\,\dot M+6\,M\,H+4\,\dot q_2\,H-4\,H^2q_2-2\,(\dot\alpha_1^2+\dot\beta_1^2)\nonumber\\
&&=-\frac{1}{2}\Delta c_0-2\,\alpha_1\left\{\frac{\partial}{\partial x}\left (p_0^2\frac{\partial H}{\partial x}\right )-\frac{\partial}{\partial y}\left (p_0^2\frac{\partial H}{\partial y}\right )\right\}\nonumber\\
&&-2\,\beta_1\left\{\frac{\partial}{\partial x}\left (p_0^2\frac{\partial H}{\partial y}\right )+\frac{\partial}{\partial y}\left (p_0^2\frac{\partial H}{\partial x}\right )\right\}\ .\label{2.31}
\end{eqnarray}
With $q_2$ given by (\ref{2.20}) we have 
\begin{eqnarray}
4\,\dot q_2\,H&-&4\,H^2q_2-2\,(\dot\alpha_1^2+\dot\beta_1^2)=-2\,(\dot\alpha_1-H\,\alpha_1)^2-2\,(\dot\beta_1-H\,\beta_1)^2\, ,\label{2.32}
\end{eqnarray}
and entering this into (\ref{2.31}) followed by multiplication across by $-1/2$ results in
\begin{eqnarray}
\dot M&-&3\,H\,M+(\dot\alpha_1-\alpha_1\,H)^2+(\dot\beta_1-\beta_1\,H)^2\nonumber \\
&=&\frac{1}{4}\Delta c_0+\alpha_1\Biggl\{\frac{\partial}{\partial x}\left (p_0^2\frac{\partial H}{\partial x}\right)-\frac{\partial}{\partial y}\left (p_0^2\frac{\partial H}{\partial y}\right )\Biggr\}\nonumber\\
&&+\beta_1\Biggl\{\frac{\partial}{\partial x}\left (p_0^2\frac{\partial H}{\partial y}\right)+\frac{\partial}{\partial y}\left (p_0^2\frac{\partial H}{\partial x}\right )\Biggr\}\ .\label{2.33}
\end{eqnarray}
To refine the estimate of $R_{AB}$ in (\ref{2.26}) we find that requiring $R_{12}=O(r^{-3})$ results in
\begin{eqnarray}
&&\frac{\partial a_3}{\partial y}+\frac{\partial b_3}{\partial x}=8\,(\dot\beta_3-3\,H\,\beta_3)-4\,m\,\beta_1-20\,\alpha_1\,\dot\alpha_1\,\beta_1 -4\,\alpha_1^2\dot\beta_1-8\,\beta_1^2\dot\beta_1-\beta_1\,\frac{\partial a_2}{\partial x}\nonumber\\
&&+8\,\beta_1\,(\beta_1^2+3\,\alpha_1^2)H-b_2\,\frac{\partial\alpha_1}{\partial x}+3\,\alpha_1\,\frac{\partial b_2}{\partial x}-3\,a_2\,\frac{\partial\beta_1}{\partial x}-\beta_1\,\frac{\partial b_2}{\partial y}+a_2\,\frac{\partial\alpha_1}{\partial y}\nonumber\\
&&-3\,\alpha_1\,\frac{\partial a_2}{\partial y}-3\,b_2\,\frac{\partial\beta_1}{\partial y}+4\,p_0^{-2}a_2\,b_2+4\,p_0^{-1}\beta_1\,\left (a_2\,\frac{\partial p_0}{\partial x}+b_2\,\frac{\partial p_0}{\partial y}\right ),\label{2.34}\end{eqnarray}
while requiring $R_{11}-R_{22}=O(r^{-3})$ results in 
\begin{eqnarray}
&&\frac{\partial a_3}{\partial x}-\frac{\partial b_3}{\partial y}=8\,(\dot\alpha_3-3\,H\,\alpha_3)-4\,m\,\alpha_1+12\,\alpha_1\,\beta_1\,\dot\beta_1+12\,\beta_1^2\dot\alpha_1-8\,\alpha_1^2\dot\alpha_1-\alpha_1\,\frac{\partial a_2}{\partial x}\nonumber \\
&&+8\,\alpha_1\,(\alpha_1^2-3\,\beta_1^2)H+b_2\,\frac{\partial\beta_1}{\partial x}-3\,\beta_1\,\frac{\partial b_2}{\partial x}-3\,a_2\,\frac{\partial\alpha_1}{\partial x}-\alpha_1\,\frac{\partial b_2}{\partial y}-a_2\,\frac{\partial\beta_1}{\partial y}\nonumber\\
&&+3\,\beta_1\,\frac{\partial a_2}{\partial y}-3\,b_2\,\frac{\partial\alpha_1}{\partial y} +2\,p_0^{-2}(a_2^2-b_2^2)+4\,p_0^{-1}\alpha_1\,\left (a_2\,\frac{\partial p_0}{\partial x}+b_2\,\frac{\partial p_0}{\partial y}\right )\ . \label{2.35}
\end{eqnarray}
The requirement that $R_{11}+R_{22}=O(r^{-3})$ provides the function $c_2(x, y, u)$ in (\ref{2.13}). This is found to be
\begin{eqnarray}
c_2&=&-2\,q_2\,c_0+p_0^{-2}(a_2^2+b_2^2)+p_0^2\left (\frac{\partial\alpha_1}{\partial x}\frac{\partial\beta_1}{\partial y}-\frac{\partial\alpha_1}{\partial y}\frac{\partial\beta_1}{\partial x}\right )\nonumber\\
&&+\frac{1}{2}p_0^6\left\{\frac{\partial}{\partial x}\left (p_0^{-4}\frac{\partial q_2}{\partial x}\right )+\frac{\partial}{\partial y}\left (p_0^{-4}\frac{\partial q_2}{\partial y}\right )\right\} \nonumber \\
&&+\frac{1}{2}p_0^2\Biggl\{\frac{\partial}{\partial x}(p_0^{-2}a_3)+\frac{\partial}{\partial y}(p_0^{-2}b_3)\Biggr\}\ .\label{2.36}
\end{eqnarray}
To complete the derivation from the vacuum field equations of equations governing the functions of $x, y, u$ which are the coefficients of the powers of $r$ in (\ref{2.8})--(\ref{2.13}), and only these functions, we find that $R_{14}=O(r^{-3})$ if
\begin{eqnarray}
&&\frac{3}{2}p_0^{-2}(\dot a_3-4\,H\,a_3)= \frac{\partial m}{\partial x}+\beta_1\,\frac{\partial c_0}{\partial y}+\alpha_1\,\frac{\partial c_0}{\partial x}\nonumber \\
&&-\frac{1}{2}\frac{\partial}{\partial y}\left\{p_0^2\left (\frac{\partial}{\partial y}(p_0^{-2}a_2)-\frac{\partial}{\partial x}(p_0^{-2}b_2)\right )\right\}\nonumber\\
&&+2\left (\dot\beta_1\,\frac{\partial\alpha_1}{\partial y}-\dot\alpha_1\,\frac{\partial\beta_1}{\partial y}\right )+\left (\beta_1\frac{\partial\dot\beta_1}{\partial x}-\dot\beta_1\frac{\partial\beta_1}{\partial x}\right )\nonumber\\
&&+\left (\alpha_1\frac{\partial\dot\alpha_1}{\partial x}-\dot\alpha_1\frac{\partial\alpha_1}{\partial x}\right )+4\,p_0^{-1}\frac{\partial p_0}{\partial y}(\dot\alpha_1\,\beta_1-\alpha_1\,\dot\beta_1)\nonumber\\
&&-\frac{1}{2}p_0^2\frac{\partial}{\partial u}\{p_0^{-4}(\alpha_1\,a_2+\beta_1\,b_2)\}\ .\label{2.37}
\end{eqnarray}
It is useful to note, using (\ref{2.22}) and (\ref{2.23}), that
\begin{eqnarray}
&&-\frac{1}{2}p_0^2\frac{\partial}{\partial u}\{p_0^{-4}(\alpha_1\,a_2+\beta_1\,b_2)\}= 2\,\frac{\partial H}{\partial x}\,q_2-\frac{1}{2}\frac{\partial\dot q_2}{\partial x}\nonumber \\
&&+2\,p_0^{-1}\frac{\partial p_0}{\partial x}(\dot q_2-2\,H\,q_2)+H\,\frac{\partial q_2}{\partial x}\nonumber\\
&&+H\left (\alpha_1\,\frac{\partial\beta_1}{\partial y}-\beta_1\,\frac{\partial\alpha_1}{\partial y}\right )+\frac{1}{2}\left (\dot\beta_1\,\frac{\partial\alpha_1}{\partial y}-\dot\alpha_1\,\frac{\partial\beta_1}{\partial y}\right )\nonumber\\
&&+\frac{1}{2}\left (\beta_1\,\frac{\partial\dot\alpha_1}{\partial y}-\alpha_1\,\frac{\partial\dot\beta_1}{\partial y}\right )\ .\label{2.38}
\end{eqnarray}
For $R_{24}=O(r^{-3})$ we must have
\begin{eqnarray}
&&\frac{3}{2}p_0^{-2}(\dot b_3-4\,H\,b_3)=\frac{\partial m}{\partial y}+\beta_1\,\frac{\partial c_0}{\partial x}-\alpha_1\,\frac{\partial c_0}{\partial y}\nonumber\\
&&-\frac{1}{2}\frac{\partial}{\partial x}\left\{p_0^2\left (\frac{\partial}{\partial x}(p_0^{-2}b_2)-\frac{\partial}{\partial y}(p_0^{-2}a_2)\right )\right\}\nonumber\\
&&-2\left (\dot\beta_1\,\frac{\partial\alpha_1}{\partial x}-\dot\alpha_1\,\frac{\partial\beta_1}{\partial x}\right )+\left (\beta_1\frac{\partial\dot\beta_1}{\partial y}-\dot\beta_1\frac{\partial\beta_1}{\partial y}\right )\nonumber\\
&&+\left (\alpha_1\frac{\partial\dot\alpha_1}{\partial y}-\dot\alpha_1\frac{\partial\alpha_1}{\partial y}\right )-4\,p_0^{-1}\frac{\partial p_0}{\partial x}(\dot\alpha_1\,\beta_1-\alpha_1\,\dot\beta_1)\nonumber\\
&&+\frac{1}{2}p_0^2\frac{\partial}{\partial u}\{p_0^{-4}(\alpha_1\,b_2-\beta_1\,a_2)\}\ ,\label{2.39}
\end{eqnarray}
and in place of (\ref{2.38}) 
\begin{eqnarray}
&&\frac{1}{2}p_0^2\frac{\partial}{\partial u}\{p_0^{-4}(\alpha_1\,b_2-\beta_1\,a_2)\}=2\,p_0^{-1}\frac{\partial p_0}{\partial y}(\dot q_2-2\,H\,q_2)\nonumber \\
&&+2\,\frac{\partial H}{\partial y}\,q_2-\frac{1}{2}\frac{\partial\dot q_2}{\partial y}+H\,\frac{\partial q_2}{\partial y}-H\left (\alpha_1\,\frac{\partial\beta_1}{\partial x}-\beta_1\,\frac{\partial\alpha_1}{\partial x}\right )\nonumber\\
&&-\frac{1}{2}\left (\dot\beta_1\,\frac{\partial\alpha_1}{\partial x}-\dot\alpha_1\,\frac{\partial\beta_1}{\partial x}\right )-\frac{1}{2}\left (\beta_1\,\frac{\partial\dot\alpha_1}{\partial x}-\alpha_1\,\frac{\partial\dot\beta_1}{\partial x}\right ).\label{2.40}
\end{eqnarray}
 
In view of the line element (\ref{2.6}) we can define a null tetrad $m^i, \bar m^i, k^i, l^i$ via the 1--forms
\begin{eqnarray}
m_i\,dx^i&=&\frac{r}{p \sqrt{2}}\Biggl\{(a+i\,b)du+e^{\alpha}(\cosh\beta+i\,\sinh\beta)dx\nonumber \\
&&+i\,e^{-\alpha}(\cosh\beta-i\,\sinh\beta)dy\Biggr\}\ ,\label{2.41}\\
\bar m_i\,dx^i&=&\frac{r}{p \sqrt{2}}\Biggl\{(a-i\,b)du+e^{\alpha}(\cosh\beta-i\,\sinh\beta)dx\nonumber \\
&&-i\,e^{-\alpha}(\cosh\beta+i\,\sinh\beta)dy\Biggr\}\ ,\label{2.42}\\
l_i\,dx^i&=&dr+\frac{1}{2}c\,du\ ,\label{2.43}\\
k_i\,dx^i&=&du\ .\label{2.44}
\end{eqnarray}
The bar denotes complex conjugation and all scalar products involving pairs of these vectors vanish with the exception of $m_i\,\bar m^i=-1$ and $l_i\,k^i=1$ confirming that (\ref{2.41})--(\ref{2.44}) constitute a null tetrad. Implementing the expansions (\ref{2.8})--(\ref{2.13}) in the calculation of the Riemann curvature tensor of the space--time we find that the leading terms in the components of the Riemann tensor on this null tetrad are given in Newman--Penrose \cite{Newman:Penrose:1962} notation by
\begin{eqnarray}
\Psi_0&=&\frac{1}{r^5}\left\{6\,(\alpha_3+i\,\beta_3)-\frac{3}{2}(\gamma+\bar\gamma)^2(\gamma-\bar\gamma)-2\bar\gamma^3\right\}+O\left (\frac{1}{r^6}\right ),\label{2.45}\\
\Psi_1&=&\frac{1}{r^4\sqrt{2}}\left\{\frac{3}{2}p_0^{-1}(a_3+i\,b_3)+3\,p_0^3\gamma\frac{\partial}{\partial\bar z}(p_0^{-2}\bar\gamma)\right\}+O\left (\frac{1}{r^5}\right ),\label{2.46}\\
\Psi_2&=&\frac{1}{r^3}\Biggl\{M+\gamma\,(\dot{\bar\gamma}-H\,\bar\gamma)+2\,p_0^2\frac{\partial}{\partial\bar z}\left (p_0^2\frac{\partial}{\partial\bar z}(p_0^{-2}\bar\gamma)\right )\Biggr\}+O\left (\frac{1}{r^4}\right ),\label{2.47}\\
\Psi_3&=&\frac{1}{r^2\sqrt{2}}\Biggl\{p_0\,\frac{\partial c_0}{\partial z}+2\,p_0\,\bar\gamma\,\frac{\partial H}{\partial\bar z}+2\,p_0^2\frac{\partial}{\partial u}\left (p_0\,\frac{\partial}{\partial\bar z}(p_0^{-2}\bar\gamma)\right )\Biggr\}+O\left (\frac{1}{r^3}\right ),\label{2.48}\\
\Psi_4&=&\frac{2}{r}\Biggl\{\frac{\partial}{\partial z}\left (p_0^2\frac{\partial H}{\partial z}\right )+\frac{1}{2}p_0^2\frac{\partial}{\partial u}\left (p_0^{-1}\frac{\partial}{\partial u}(p_0^{-1}\bar\gamma)\right )\Biggr\}+O\left (\frac{1}{r^2}\right )\ ,\label{2.49}
\end{eqnarray}
with $\gamma=\alpha_1+i\,\beta_1$ and $z=x+i\,y$. We have here a manifestation of the ``peeling theorem'' of Sachs \cite{Sachs:1962}{}.

\section{Model of an Accelerating Massive Particle}\label{sec:3}   
    
As a preliminary step in constructing our model particle we specialize the line element (\ref{2.6}) to a form of the line element of Minkowskian space--time suitable for our purposes. This form of the Minkowskian line element is 
\begin{equation}\label{3.1}
ds^2=-r^2P_0^{-2}(dx^2+dy^2)+2\,du\,dr+(1-2\,\underset{0}{H}\,r)du^2\ ,
\end{equation}
with
\begin{eqnarray}
&&P_0=x\,v^1(u)+y\,v^2(u)+\left (1-\frac{1}{4}(x^2+y^2)\right )v^3(u)+\left (1+\frac{1}{4}(x^2+y^2)\right )v^4(u)\ , \nonumber\\\label{3.2}\\
&&-(v^1)^2-(v^2)^2-(v^3)^2+(v^4)^2=1\ ,\label{3.3}
\end{eqnarray}
and
\begin{equation}\label{3.4}
\underset{0}{H}=\frac{\partial}{\partial u}(\log P_0)\ .
\end{equation}
This corresponds to (\ref{2.6}) with $\alpha=\beta=a=b=0$, $p_0=P_0$ and $c=1-2\,\underset{0}{H}r$ ($\Rightarrow\ c_{-1}=-2\,H$ with $H=\underset{0}{H}$ and $c_0=1$ in (\ref{2.13})). We can write (\ref{3.1}) as $ds^2=\eta_{ij}\,dX^i\,dX^j$ with $\eta_{ij}={\rm diag}(-1, -1, -1, +1)$ and the rectangular Cartesian and time coordinates $X^i$ are related to the coordinates $x, y, r, u$ by
\begin{eqnarray}
X^i&=&w^i(u)+r\,k^i\ \ {\rm with}\ \label{3.5} \\
k^i&=&P_0^{-1}\left (-x, -y, -1+\frac{1}{4}(x^2+y^2), 1+\frac{1}{4}(x^2+y^2)\right )\ . \nonumber
\end{eqnarray}
Also $r=0$ is an arbitrary time--like world line $X^i=w^i(u)$ with unit time--like tangent $v^i(u)=dw^i/du$ and $\eta_{ij}\,v^i\,v^j=v_j\,v^j=1$ as in (\ref{3.3}). Hence $v^i(u)$ is the 4--velocity of the particle with world line $r=0$ and $u$ is proper--time along the world line. The 4--acceleration is $a^i(u)=dv^i/du$ and as a consequence of (\ref{3.3}) $\eta_{ij}\,v^i\,a^j=v_j\,a^j=0$. With $P_0$ given by (\ref{3.2}) and $k^i$ by (\ref{3.5}) we can write $\underset{0}{H}$ in (\ref{3.4}) as $\underset{0}{H}=\eta_{ij}\,a^i\,k^j=a_j\,k^j$. The vector field $k^i$ is light--like and normalized so that 
\begin{equation}\label{3.6}
\eta_{ij}\,k^i\,k^j=k_j\,k^j=0\ \ \ {\rm and}\ \ \ \eta_{ij}\,k^i\,v^j=k_j\,v^j=1\ .
\end{equation}

\begin{figure}[pb]
    \centerline{\includegraphics[width=0.5\textwidth]{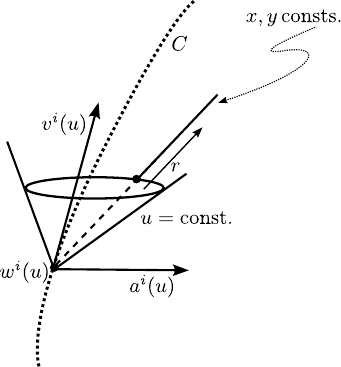}}
    \vspace*{8pt}
    \caption{\label{fig_2} In the Minkowskian background space--time $C$ is the time--like world line $r=0$ of the accelerating mass particle and $u={\rm constant}$ are the future null cone histories of the radiation. The gravitational field of the particle is a tensor field on Minkowskian space--time which is singular on $r=0$ and also on generators of the null cones labelled by $x=\pm\infty$ and/or $y=\pm\infty$.}
\end{figure}

An accelerating charged particle produces electromagnetic radiation which, as a consequence of Maxwell's field equations, is shear--free. The charge on the particle is conserved but the particle experiences radiation reaction. We consider an approximate model initiated in Ref.\ \refcite{Hogan:Robinson:1986} of an accelerating mass particle producing gravitational radiation which is shearing. The particle experiences radiation reaction similar to the charged particle but its mass is not conserved and diminishes due to the shear in the radiation. The field of the particle is an approximate solution of Einstein's vacuum field equations and the spacetime is a perturbation of Minkowskian spacetime with the perturbation singular on the world line of the mass particle in Minkowskian spacetime. Our purpose here is to develop a model which illustrates in the gravitational context some of the qualitative properties of an accelerating charge in electromagnetic theory. In terms of the variables introduced in the previous section we begin by making the assumptions that
\begin{equation}\label{3.7}
\alpha_1=\alpha_1(u)\ ,\ \beta_1=\beta_1(u)\ ,
\end{equation}
so that $\alpha_1, \beta_1$ are independent of the coordinates $x, y$, and in addition we simplify (\ref{2.30}) by taking
\begin{equation}\label{3.8}
m(x, y, u)=\frac{1}{2}\,p_0^2\left\{\frac{\partial}{\partial x}(p_0^{-2}a_2)+\frac{\partial}{\partial y}(p_0^{-2}b_2)\right\}+m_0(u)\ ,
\end{equation}
with $m_0$ some function of $u$. Next we introduce approximations by writing
\begin{equation}\label{3.9}
p_0=P_0\,(1+Q)\ \ {\rm with}\ \ Q=O_1\ ,
\end{equation}
meaning that $Q(x, y, u)$ is small of first order and $P_0$ is given by (\ref{3.2}). We also assume that $m_0(u)$ in (\ref{3.8}) is small of first order (so that we write $m_0=O_1$) and $\alpha_1, \beta_1$ are both small of order one half (so that $\alpha_1^2=O_1$ and $\beta_1^2=O_1$). To make clear the meaning of these quantities being small we first note that with our units for which $c=G=1$ we have $Q$ dimensionless and $m_0$, $\alpha_1$ and $\beta_1$ each have dimensions of length. Hence with $a^i(u)=dv^i/du$ the 4--acceleration of the world line of the mass particle in the background Minkowskian spacetime (and thus $a^i$ has the dimensions of inverse length) we have in mind that $m_0\times(-a_i\,a^i)^{1/2}=O_1$ is dimensionless and small of first order while $\alpha_1\times(a_i\,a^i)^{1/2}=O_{1/2}$ and $\beta_1\times(a_i\,a^i)^{1/2}=O_{1/2}$ are both dimensionless and small of order one half. Consequently we write, following from (\ref{2.20}), $q_2=O_1$ and from (\ref{2.16}), (\ref{2.22}) and (\ref{2.23}),
\begin{eqnarray}
c_{-1}&=&-2\,H\ \ \ {\rm with}\quad H=\underset{0}{H}+\dot Q+O_2\ \, {\rm and}\ \ \underset{0}{H}=P_0^{-1}\dot P_0=a_i\,k^i,\label{3.10}\\
c_0&=&1+\underset{0}{\Delta} Q+2\,Q+O_2\ \, {\rm with}\ \ \ \underset{0}{\Delta}=P_0^2\left (\frac{\partial^2}{\partial x^2}+\frac{\partial ^2}{\partial y^2}\right )\ ,\label{3.11}\\
a_2&=&P_0^4\left\{\frac{\partial}{\partial x}(P_0^{-2}\alpha_1)+\frac{\partial}{\partial y}(P_0^{-2}\beta_1)\right\}+O_{3/2}\ \nonumber\\
&=&-2\,P_0\frac{\partial P_0}{\partial x}\,\alpha_1-2\,P_0\frac{\partial P_0}{\partial y}\,\beta_1+O_{3/2}\ ,\label{3.12}\\
b_2&=&P_0^4\left\{\frac{\partial}{\partial x}(P_0^{-2}\beta_1)-\frac{\partial}{\partial y}(P_0^{-2}\alpha_1)\right\}+O_{3/2}\ \nonumber\\
&=&2\,P_0\frac{\partial P_0}{\partial y}\,\alpha_1-2\,P_0\frac{\partial P_0}{\partial x}\,\beta_1+O_{3/2}\ ,\label{3.13}
\end{eqnarray}
Here $\underset{0}{\Delta}$ is the Laplacian on the unit sphere (so that if, for example, $F(x, y, u)$ is a spherical harmonic of order $l$ then $\underset{0}{\Delta}F+l\,(l+1)\,F=0$ for $l=0, 1, 2, \dots$). With $m(x, y, u)$ given by (\ref{3.8}) and the approximations introduced above we can write (\ref{2.30}) as
\begin{equation}\label{3.14}
M=m_0(u)-\alpha_1\,\dot\alpha_1-\beta_1\,\dot\beta_1+(\alpha_1^2+\beta_1^2)\underset{0}{H}+O_{3/2}=O_1\ .
\end{equation}
From now on we will consistently neglect small terms of order three-halves and so we will consider the symbol $O_{3/2}$ to be understood. Substituting the approximations into the field equation (\ref{2.33}) we can write the result in the useful form
\begin{eqnarray}
\frac{1}{4}\underset{0}{\Delta}(\underset{0}{\Delta} Q+2\,Q)&=&\dot m_0-\alpha_1\,\ddot\alpha_1-\beta_1\,\ddot\beta_1+3(\alpha_1\,\dot\alpha_1+\beta_1\,\dot\beta_1-m_0)\underset{0}{H}\nonumber\\
&&+(\alpha_1^2+\beta_1^2)(\underset{0}{\dot H}+\underset{0}{H^2}+a_i\,a^i)-3(\alpha_1^2+\beta_1^2)(\underset{0}{H^2}+\frac{1}{3}a_i\,a^i).\label{3.15}
\end{eqnarray}
The first line on the right hand side of this equation is an $l=0$ spherical harmonic. The second and the third lines on the right hand side are $l=1$ spherical harmonic and the last line is an $l=2$ spherical harmonic. The histories of the wave fronts produced by the accelerating particle are the null hypersurfaces $u={\rm constant}$ and we expect the wave fronts, corresponding to $u={\rm constant}$ and $r={\rm constant}$, to be smoothly deformed 2--spheres. This means that the function $Q(x, y, u)$ should be a well--behaved function of the stereographic coordinates $x, y$ for $-\infty< x <+\infty$ and $-\infty< y <+\infty$. To achieve this we must first put to zero the $l=0$ spherical harmonic on the right hand side of (\ref{3.15}) giving us the equation
\begin{equation}\label{3.16}
\dot m_0=\alpha_1\,\ddot\alpha_1+\beta_1\,\ddot\beta_1\ .
\end{equation}
Now we can solve (\ref{3.15}) with
\begin{eqnarray}
\frac{1}{4}(\underset{0}{\Delta} Q+2\,Q)&=&-\frac{3}{2}(\alpha_1\,\dot\alpha_1+\beta_1\,\dot\beta_1-m_0)\underset{0}{H}-\frac{1}{2}(\alpha_1^2+\beta_1^2)(\underset{0}{\dot H}+\underset{0}{H^2}+a_i\,a^i)\nonumber\\
&&+\frac{1}{2}(\alpha_1^2+\beta_1^2)(\underset{0}{H^2}+\frac{1}{3}a_i\,a^i)\ .\label{3.17}
\end{eqnarray}
It is helpful to rewrite the first two lines on the right hand side here using
\begin{equation}\label{3.18}
\underset{0}{H}=a_i\,k^i\ \ \Rightarrow\ \ \underset{0}{\dot H}+\underset{0}{H^2}=\dot a_i\,k^i\ ,
\end{equation}
since 
\begin{equation}\label{3.19}
\frac{\partial k^i}{\partial u}=-\underset{0}{H}\,k^i\ .
\end{equation}
Next write
\begin{equation}\label{3.20}
k^i=v^i+p^i\ \ {\rm with}\ \ p^i\,v_i=0\ \ {\rm and}\ \ p^i\,p_i=-1\ .
\end{equation}
The arbitrary spatial direction of the null vector field $k^i$ (the projection of $k^i$ orthogonal to $v^i$) is given by the unit space--like vector $p^i$. Now
\begin{equation}\label{3.21}
\underset{0}{\dot H}+\underset{0}{H^2}=\dot a_i\,k^i=\dot a_j\,h^j_ip^i-a_i\,a^i\ ,
\end{equation}
with $h^i_j=\delta^i_j-v^i\,v_j$ the projection tensor projecting vectors orthogonal to $v^i$. For $Q$ in (\ref{3.17}) to be a well--behaved function of $x, y$ the $l=1$ spherical harmonic on the right hand side must vanish. This condition can be written
\begin{equation}\label{3.22}
3(\alpha_1\,\dot\alpha_1+\beta_1\,\dot\beta_1-m_0)a_i\,p^i+(\alpha_1^2+\beta_1^2)\dot a_j\,h^j_i\,p^i=0\ .
\end{equation}
This equation must hold for all unit vectors $p^i$ orthogonal to $v^i$ and so it results in 
\begin{equation}\label{3.23}
{\cal M}\,a^i=\frac{1}{3}(\alpha_1^2+\beta_1^2)h^i_j\dot a_j=\frac{1}{3}|\gamma|^2(\dot a^i+a_j\,a^j\,v^i)\ ,
\end{equation}
with
\begin{equation}\label{3.24}
{\cal M}(u)=m_0-\alpha_1\,\dot\alpha_1-\beta_1\,\dot\beta_1\ .
\end{equation}
Now using (\ref{3.16}) and (\ref{3.24}) we arrive at
\begin{equation}\label{3.25}
\dot{\cal M}=-(\dot\alpha_1^2+\dot\beta_1^2)=-|\dot\gamma|^2\ ,
\end{equation}
remembering that $\gamma$ is defined following (\ref{2.49}). In eq.(\ref{3.23}) we have derived an approximate equation of motion for our model massive particle. The terms on both sides of the equation are dimensionless and small of first order and we have neglected terms small of order three--halves. It appears qualitatively similar to a Lorentz--Dirac equation of motion of a charged particle in electrodynamics. The appearance of ${\cal M}$ on the left hand side of (\ref{3.23}) suggests that we should take ${\cal M}$ to be the mass of the particle. There is a mass--loss formula given by (\ref{3.25}) due to the variable shear in the gravitational radiation emitted by the particle and described below. The quantity $|\dot\gamma|$ here is playing the role of Bondi's \emph{news function} \cite{Bondi:etal:1962} and if ${\cal M}$ is taken as the mass of the particle then (\ref{3.25}) is a special case of Bondi's statement that ``the mass of a system is constant if and only if there is no news. If there is news, the mass decreases monotonically as long as the news continues''. The two equations (\ref{3.23}) and (\ref{3.25}) can be written together in the form
\begin{equation}\label{3.26}
\frac{d}{du}({\cal M}\,v^i)=-|\dot\gamma|^2v^i+\frac{1}{3}|\gamma|^2h^i_j\,\dot a^j\ .
\end{equation}
Returning now to (\ref{3.17}), with the first two lines on the right hand side put to zero, we obtain for $Q$ the $l=2$ spherical harmonic
\begin{equation}\label{3.27}
Q(x, y, u)=-\frac{1}{2}\,|\gamma|^2\left (\underset{0}{H^2}+\frac{1}{3}\,a_i\,a^i\right )=O_1\ .
\end{equation}
We note that an $l=0$ or $l=1$ term in $Q$ corresponds to a trivial perturbation of the otherwise spherical wave fronts, so that they remain spherical. When viewed against a Euclidean 3--space background such terms result either in an infinitesimal perturbation of the radius of a 2-sphere (if $l=0$) or in an infinitesimal displacement of the centre of a 2--sphere (if $l=1$).
 
\section{The Gravitational Field of the Particle}\label{sec:4}

When the approximations described in section \ref{sec:3} are introduced into the Newman--Penrose components (\ref{2.45})--(\ref{2.49}) of the Riemann curvature tensor (the gravitational field of the particle) we obtain
\begin{eqnarray}
\Psi_0&=&-\frac{6}{r^5}\,\{\alpha_3+i\,\beta_3+O_{3/2}\}+O\left (\frac{1}{r^6}\right )\ ,\label{4.1}\\
\Psi_1&=&-\frac{3}{r^42\,\sqrt{2}}\left\{P_0^{-1}(a_3+i\,b_3)-4\,|\gamma|^2\frac{\partial P_0}{\partial z}+O_{3/2}\right\}+O\left (\frac{1}{r^5}\right )\ ,\label{4.2}\\
\Psi_2&=&-\frac{1}{r^3}\left\{P_0^2\frac{\partial}{\partial\bar z}(P_0^{-2}(a_2-i\,b_2))+{\cal M}+\dot{\bar\gamma}\,\gamma+O_{3/2}\right\}+O\left (\frac{1}{r^4}\right )\ ,\label{4.3}\\
\Psi_3&=&\frac{\sqrt{2}}{r^2}\Biggl\{\bar\gamma\,P_0\frac{\partial }{\partial\bar z}(\underset{0}{H})+2\,(\dot{\bar\gamma}-\underset{0}{H}\,\bar\gamma)\frac{\partial P_0}{\partial\bar z}-2\,|\gamma|^2P_0\,\underset{0}{H}\,\frac{\partial}{\partial z}(\underset{0}{H})+O_{3/2}\Biggr\}+O\left (\frac{1}{r^3}\right )\ ,\nonumber \\ \label{4.4}\\
\Psi_4&=&\frac{1}{r}\Biggl\{-P_0^2\frac{\partial}{\partial u}\left (P_0^{-2}(\dot{\bar\gamma}-\underset{0}{H}\,\bar\gamma)\right )+2\left (\frac{\partial}{\partial u}-2\,\underset{0}{H}\right)|\gamma|^2P_0^2\left (\frac{\partial\underset{0}{H}}{\partial z}\right )^2+O_{3/2}\Biggr\}\nonumber \\
&&+O\left (\frac{1}{r^2}\right )\ .\label{4.5}
\end{eqnarray}
The functions $\alpha_3, \beta_3, a_3, b_3$ of $x, y, u$ appearing in (\ref{4.1}) and (\ref{4.2}) are obtained by specializing, with the approximations introduced in section \ref{sec:3}, the field equations (\ref{2.34}), (\ref{2.35}) and (\ref{2.37}), (\ref{2.39}). The latter two equations involve the function $m(x, y, u)$ given now by (\ref{3.8}). With the approximations of section \ref{sec:3}, (\ref{3.8}) reads
\begin{eqnarray}
m(x, y, u)&=&\left\{\left (\frac{\partial P_0}{\partial x}\right )^2-\left (\frac{\partial P_0}{\partial y}\right )^2\right\}\alpha_1+2\,\frac{\partial P_0}{\partial x}\,\frac{\partial P_0}{\partial y}\,\beta_1+m_0(u)+O_{3/2}\ .\label{4.6}
\end{eqnarray}
We also find that
\begin{eqnarray}
&&p_0^2\left (\frac{\partial}{\partial y}(p_0^{-2}a_2)-\frac{\partial}{\partial x}(p_0^{-2}b_2)\right )=4\,\frac{\partial P_0}{\partial x}\frac{\partial P_0}{\partial y}\alpha_1\nonumber \\
&&-2\Biggl\{\left (\frac{\partial P_0}{\partial x}\right )^2-\left (\frac{\partial P_0}{\partial y}\right )^2\Biggr\}\beta_1+O_{3/2}\ .\label{4.7}
\end{eqnarray} 
In preparation for substitution into (\ref{2.37}) we have from (\ref{4.6}) and (\ref{4.7}) that
\begin{equation}\label{4.8}
\frac{\partial m}{\partial x}-\frac{1}{2}\frac{\partial}{\partial y}\left\{p_0^2\left (\frac{\partial}{\partial y}(p_0^{-2}a_2)-\frac{\partial}{\partial x}(p_0^{-2}b_2)\right )\right\}=O_{3/2}\ .
\end{equation}
In view of (\ref{3.7}) and (\ref{4.8}) we obtain from (\ref{2.37}) with (\ref{2.38}) the approximate equation to be satisfied by $a_3(x, y, u)=O_1$:
\begin{eqnarray}
\dot a_3-4\,\underset{0}{H}\,a_3&=&\frac{4}{3}P_0\frac{\partial P_0}{\partial x}(\dot q_2-2\underset{0}{H}\,q_2)+\frac{4}{3}P_0^2\frac{\partial\underset{0}{H}}{\partial x}q_2\nonumber\\
&&+\frac{8}{3}P_0\frac{\partial P_0}{\partial y}(\dot\alpha_1\,\beta_1-\alpha_1\,\dot\beta_1)+O_{3/2}\ .\label{4.9}
\end{eqnarray}
A similar specialization of (\ref{2.39}) with (\ref{2.40}) yields the approximate equation to be satisfied by $b_3(x, y, u)=O_1$:
\begin{eqnarray}
\dot b_3-4\,\underset{0}{H}\,b_3&=&\frac{4}{3}\,P_0\frac{\partial P_0}{\partial y}(\dot q_2-2\underset{0}{H}\,q_2)+\frac{4}{3}\,P_0^2\frac{\partial\underset{0}{H}}{\partial y}q_2
\nonumber\\ 
&&-\frac{8}{3}P_0\frac{\partial P_0}{\partial x}(\dot\alpha_1\,\beta_1-\alpha_1\,\dot\beta_1)+O_{3/2}\ .\label{4.10}
\end{eqnarray}
When the approximations of section \ref{sec:3} are introduced into (\ref{2.34}) and (\ref{2.35}) we arrive at the following equations to be satisfied by $\alpha_3(x, y, u)=O_1$ and $\beta_3(x, y, u)=O_1$:
\begin{eqnarray}
8\,(\dot\alpha_3-3\,\underset{0}{H}\,\alpha_3)&=&4\,\left\{\left (\frac{\partial P_0}{\partial x}\right )^2-\left (\frac{\partial P_0}{\partial y}\right )^2\right\}\,q_2+\frac{\partial a_3}{\partial x}-\frac{\partial b_3}{\partial y}+O_{3/2}\ ,\label{4.11}
\end{eqnarray}
and
\begin{equation}\label{4.12}
8\,(\dot\beta_3-3\,\underset{0}{H}\,\beta_3)=\frac{\partial a_3}{\partial y}+\frac{\partial b_3}{\partial x}+8\,\frac{\partial P_0}{\partial x}\,\frac{\partial P_0}{\partial y}\,q_2+O_{3/2}\ .
\end{equation}
If non--zero (\ref{4.5}) represents the part of the gravitational field due to the presence of gravitational radiation. In general if there is non--zero \emph{news} and if the mass particle is accelerating then gravitational radiation exists. We see that if the mass particle has zero 4--acceleration ($a^i=0$) but non--vanishing \emph{news} ($|\dot\gamma|\neq 0$) then gravitational radiation is present since
\begin{equation}\label{4.13}
\Psi_4=-\frac{1}{r}\{\ddot{\bar\gamma}+O_{3/2}\}+\dots\ .
\end{equation}
This replicates a corresponding result of Bondi \cite{Bondi:etal:1962} (his eqn.(45)). On the other hand if $a^i\neq 0$, $\gamma\neq 0$ and $\dot\gamma=0$, so there is no \emph{news}, then gravitational radiation with shear is present since
\begin{equation}\label{4.14}
\Psi_4=\frac{1}{r}\{(\underset{0}{\dot H}-2\,\underset{0}{H}^2)\bar\gamma+O_1\}+\dots\ .
\end{equation}
There is no mass loss (in the sense that $\dot{\cal M}=0$) in this case.

We note that we can rewrite (\ref{4.11}) and (\ref{4.12}) in the form
\begin{equation}\label{4.15}
\dot a'_3-4\,\underset{0}{H}\,a'_3=\frac{8}{3}P_0\,\frac{\partial P_0}{\partial y}(\dot\alpha_1\beta_1-\alpha_1\dot\beta_1)\ ,
\end{equation}
and
\begin{equation}\label{4.16}
\dot b'_3-4\,\underset{0}{H}\,b'_3=-\frac{8}{3}P_0\,\frac{\partial P_0}{\partial x}(\dot\alpha_1\beta_1-\alpha_1\dot\beta_1)\ ,
\end{equation}
respectively, with
\begin{equation}\label{4.17}
a'_3=a_3-\frac{4}{3}P_0\frac{\partial P_0}{\partial x}\,q_2\ \ \ {\rm and}\ \ \ b'_3=b_3-\frac{4}{3}P_0\frac{\partial P_0}{\partial y}\,q_2\ .
\end{equation}
Now (\ref{4.11}) and (\ref{4.12}) can be written
\begin{eqnarray}
\dot\alpha_3-3\,\underset{0}{H}\alpha_3&=&\frac{2}{3}\left\{\left (\frac{\partial P_0}{\partial x}\right )^2-\left (\frac{\partial P_0}{\partial y}\right )^2\right\}q_2+\frac{1}{8}\left (\frac{\partial a'_3}{\partial x}-\frac{\partial b'_3}{\partial y}\right )+O_{3/2}\ ,\label{4.18}
\end{eqnarray}
and
\begin{equation}\label{4.19}
\dot\beta_3-3\,\underset{0}{H}\beta_3=\frac{1}{8}\left (\frac{\partial a'_3}{\partial y}+\frac{\partial b'_3}{\partial x}\right )+\frac{4}{3}\frac{\partial P_0}{\partial x}\frac{\partial P_0}{\partial y}q_2+O_{3/2}\ .
\end{equation}
It follows from these equations, on account of the explicit dependence of $P_0$ in (\ref{3.2}) on the stereographic coordinates $(x, y)$, that the functions $\alpha_3, \beta_3, a_3, b_3$ are singular when $x\rightarrow\pm\infty$ and/or $y\rightarrow\pm\infty$. Specifying values of $(x, y)$ selects a generator on each of the null hypersurfaces $u={\rm constant}$ (which are future null cones in the Minkowskian background and in the perturbed space--time for small values of $r$). All of the components (\ref{4.1})--(\ref{4.5}) of the Riemann tensor display this singular behavior. This is to be expected since there is no external field to supply energy to the particle. The existence of this type of singularity is known in Robinson--Trautman \cite{Robinson:Trautman:1960} fields and the observation of P.G.\ Bergmann quoted in Ref.\ \refcite{Robinson:Trautman:1960} applies equally here, namely, that such singularities ``might conceivably represent a flow of matter which restores to the source the energy carried away by radiation''.

\section{Discussion}\label{sec:5}

If the gravitational radiation produced via the acceleration and the mass loss of the massive particle described in sections \ref{sec:3} and \ref{sec:4} is shear--free then, as far as we have carried out the calculations, we see from (\ref{2.15}) that $\alpha_1=\beta_1=\alpha_3=\beta_3=0$ (and thus in particular $\gamma=0$). Consequently (\ref{3.23}) implies $a^i=0$ so that the world line of the particle in the background Minkowskian space--time is a time--like geodesic and, from (\ref{3.25}), that ${\cal M}={\rm constant}$ and by (\ref{3.8}) and (\ref{3.24}) ${\cal M}=m_0=m$ now. Also $a_3, b_3$ are, approximately, functions of $x, y$ only (i.e.\ independent of $u$) on account of (\ref{4.9}) and (\ref{4.10}). It follows from (\ref{4.11}) and (\ref{4.12}) that $a_3, b_3$ satisfy the Cauchy--Riemann equations and so can be transformed away (see eqs.\ (3.6) and (3.7) in Ref.\ \refcite{Hogan:Trautman:1987}). Now all the leading terms (\ref{4.1})--(\ref{4.5}) in the curvature tensor vanish with the exception of $\Psi_2=-m/r^3+O(r^{-4})$. With $a^i=0$ we can take $v^i=\delta^i_4$ and then, by (\ref{3.2}), $P_0=1+\frac{1}{4}(x^2+y^2)$ and $Q=0$ in (\ref{3.27}). Hence we have arrived in this case at an approximate version of the Schwarzschild solution of Einstein's vacuum field equations. In the notation of section \ref{sec:3} the exact Schwarzschild line element can be written
\begin{eqnarray}
ds^2=-r^2P_0^{-2}(dx^2+dy^2)+2\,du\,dr+\left (1-\frac{2\,m}{r}\right )du^2\ ,\nonumber
\end{eqnarray}
with $P_0$ given by (\ref{3.2}) and $a^i=dv^i/du=0$ so that the world line $r=0$ in the background Minkowskian space--time is an arbitrary time--like geodesic. Unlike this particle however, the particle considered in section 3 is accelerating with equations of motion (\ref{3.24}) and is thus radiating gravitationally as indicated in (\ref{4.14}).

A remaining challenge is to include an external gravitational field which should appear in the interesting equation (\ref{3.23}) in the form of an external 4--force and in the curvature tensor components (\ref{4.1})--(\ref{4.5}) to counter the singularities there when the stereographic coordinates become infinite. 

\bibliographystyle{ws-ijmpd}
\bibliography{opticalshear_bibliography}
\end{document}